\documentclass[fleqn,10pt]{wlscirep}

\usepackage{epstopdf}
\usepackage{bm}

\graphicspath{{fig/}{./fig/}{.}}

\title{Magnetic Lifshitz transition and its consequences in multi-band iron-based superconductors}

\author[1,2,*]{Andrzej Ptok}
\author[3]{Konrad J. Kapcia}
\author[4]{Agnieszka Cichy}
\author[5,6]{Andrzej M. Ole\'{s}}
\author[2]{Przemys\l{}aw Piekarz}
\affil[1]{Institute of Physics, Maria Curie-Sk\l{}odowska University, Plac M. Sk\l{}odowskiej-Curie 1, PL-20031 Lublin, Poland}
\affil[2]{Institute of Nuclear Physics, Polish Academy of Sciences, ul. E. Radzikowskiego 152, PL-31342 Krak\'ow, Poland}
\affil[3]{Institute of Physics, Polish Academy of Sciences, Aleja Lotnik\'{o}w 32/46, PL-02668 Warsaw, Poland}
\affil[4]{Institut f\"{u}r Physik, Johannes Gutenberg-Universit\"{a}t Mainz, Staudingerweg 9, D-55099 Mainz, Germany}
\affil[5]{Marian Smoluchowski Institute of Physics, Jagiellonian University, ul. prof. S. \L{}ojasiewicza 11, PL-30348 Krak\'ow, Poland}
\affil[6]{Max Planck Institute for Solid State Research, Heisenbergstrasse 1, D-70569 Stuttgart, Germany}

\affil[*]{aptok@mmj.pl (A.P.), konrad.kapcia@ifpan.edu.pl (K.J.K.), agnieszkakujawa2311@gmail.com (A.C.), amoles@fkf.mpg.de (A.M.O.), piekarz@wolf.ifj.edu.pl (P.P.)}

\keywords{Lifshitz transition, BCS-BEC crossover, iron-based superconductors}

\begin{abstract}
In this paper we address Lifshitz transition induced by applied
external magnetic field in a case of iron-based superconductors,
in which a difference between the Fermi level and the edges of the
bands is relatively small. We introduce and investigate a two-band
model with intra-band pairing in the relevant parameters regime to
address a generic behaviour of a system with hole-like and
electron-like bands in external magnetic field. Our results show that
two Lifshitz transitions can develop in analysed systems and the first
one occurs in the superconducting phase and takes place at
approximately constant magnetic field. The chosen sets of the model
parameters can describe characteristic band structure of iron-based
superconductors and thus the obtained results can explain the
experimental observations in FeSe and Co-doped BaFe$_{2}$As$_{2}$
compounds.
\end{abstract}
\begin{document}

\flushbottom
\maketitle

%
\thispagestyle{empty}

\section*{Introduction}

The Lifshitz transition (LT) is an electronic topological transition~\cite{lifshitz.60}.
A consequence of this transition is a change of the Fermi surface (FS) topology of a metal due to the variation of the Fermi energy and/or the band structure.
The LT can by induced by external pressure, doping or external magnetic field and has been experimentally observed in many real systems, such as e.g. heavy-fermion systems~\cite{daou.bergemann.06,bercx.assaad.12,gertrud,aoki.seyfarth.16},
iron-based superconductors~\cite{liu.kondo.10,nakayma.sato.11,malaeb.shimojima.12,xu.richard.13,khan.duane.14,liu.lograsso.14,Rod16}, cuprate high-temperature superconductors~\cite{norman.lin.10,leboeuf.doiron.11,benhabib.sacuto.15}, or other strongly correlated electrons system~\cite{okamoto.nishio.10}.

In particular, the LT induced by doping has been found in iron-based superconductors.
This family of materials is characterized by the existence of FeX layers in their crystal structure (where X is As, P, S, Se or Te).
The consequence of this feature is a specific band structure as well as the occurrence of FSs created by hole- and electron-like pockets near the $\Gamma$ and $M$ points of the first Brillouin zone (FBZ), respectively.
In general, these systems are very sensitive to doping~\cite{kordyuk.12} and the LT induced by doping can be observed e.g. in angle-resolved photoemission spectroscopy (ARPES) experiments.
For example, in the Ba$_{1-x}$K$_{x}$FeAs compounds~\cite{nakayma.sato.11,malaeb.shimojima.12,xu.richard.13,khan.duane.14,liu.lograsso.14}, a partial vanishing of the FS at the $M$ point can be observed as well as changes of its shape with potassium doping.
Other possibility is a realisation of the LT induced by magnetic filed, which we will call {\it the magnetic Lifshitz transition} (MLT) below.
The MLT in multi-band systems is interesting in the case of relatively {\it small} FS pockets, observed e.g. in FeSe~\cite{kasahara.watashige.14}, where relatively large external magnetic field can lead to disappearance of the FS for one of spin types.
This situation has been also discussed widely in the context of heavy-fermion systems, e.g. in CeIn$_{3}$, where the reconstruction of the FS inside the N\'{e}el long-range order phase has been observed~\cite{harrison.sebastian.07}.
In this case, the {\it small} FS pockets collapse in relatively large external magnetic field and become depopulated~\cite{schlottmann.11}, while the changes in the FS topology occur at approximately constant value of the magnetic field.
In CeRu$_{2}$Si$_{2}$ the behaviour is different, where only one-spin FS pocket disappears~\cite{daou.bergemann.06}.
As we can see, one can expect a realisation of two types of the MLT: {\it entire} or {\it partial}.
For the former, as in a typical LT induced by doping, the magnetic field changes the band structure for electrons with both spins, while in the latter only one type of the electron spin band (FS) disappears at the Fermi level.
In addition, this type of the LT can be important in the case where a {\it small} FS exists and it is very sensitive to the magnetic field~\cite{daou.bergemann.06}.

The iron based superconductors belong to systems with strong electron 
correlations~\cite{dagotto.13,Massimo}.
In such systems, the unconventional superconductivity with a non-trivial Cooper pairing lay down one of the most important directions of studies in the theory of condensed matter and ultracold quantum gases.
There are indications that the properties of unconventional superconductors place them between two regimes: BCS and BEC~\cite{robaszkiewicz.micnas.81b,robaszkiewicz.micnas.82,robaszkiewicz.micnas.87,
domanski.ranninger.01,domasnki.ranninger.04,chen.stajic.05,micnas.ranninger.90,kapcia.robaszkiewicz.12,kapcia.robaszkiewicz.13}.
In a case of a single-band attractive Hubbard model the tightly bound local pairs of fermions behave as hard-core bosons and they can exhibit a superfluid state similar to that of $^{4}\textrm{He}$ II.
The evolution from the weak attraction (BCS-like) to that of the strong attraction (BEC-like) limit takes place when the interaction is increased or the electron concentration is decreased at moderate fixed attraction~\cite{micnas.ranninger.90,kapcia.robaszkiewicz.12,kapcia.robaszkiewicz.13}.
According to the {\it Leggett criterion}~\cite{leggett.80}, the Bose regime begins when the chemical potential $\mu$ drops below the lower band edge.
Strong renormalization of the chemical potential $\mu$ and the possibility of the complete condensation of all the electrons of the sub-band in which the chemical potential is close to its energy bottom, give rise to the coexistence of BCS-like and BEC-like pairs~\cite{bianconi.valletta.98}.
However, formally BCS-BEC crossover regime exist, when the size of interacting pairs (coherence length $\zeta$) becomes comparable to the average distance between particles ($\approx 1/k_{F}$), i.e. $k_{F}\zeta \approx 1$, where $k_{F}$ is a the Fermi momentum.
In this regime, the character of a Fermi superfluid continuously change from the weak-coupling BCS type to the BEC of tightly bound molecules with increasing the strength interaction, and there is no distinct phase boundary between the weak-coupling BCS regime and the strong-coupling BEC regime.
The BCS to BEC evolution in a two-band system is much richer than the one-band case and have been discussed in several papers in the context of two electron-like bands~\cite{iskin.sademelo.06,iskin.sademelo.07,guidini.perali.14,guidini.flammia.16}.
Also, in iron based superconductors BCS-BEC crossover regime has been reported, e.g. FeSe~\cite{kasahara.watashige.14}.
In this case, a superconducting gap is comparable to the Fermi level E$_{F}$, e.g. in FeSe compound, the ratio between gap and the Fermi level are estimated as $1$ ($0.3$) in the electron- (hole-) like band~\cite{kasahara.watashige.14}, while in Te-doped FeSe as $0.5$ in hole band~\cite{lubashevsky.lahoud.12}.
Because the gap energy is comparable with the Fermi level of the electron band, the system can be treated as a material in the BCS-BEC crossover regime~\cite{chubukov.eremin.16}.
This property is a generic feature of all iron-based superconductors~\cite{bianconi.13},
with the consequence that the transition from the BCS to the BEC limit
can have important influence on the value of T$_{c}$~\cite{kagan.16}.

The recent results report an additional phase transition in a highly polarized superconducting state phase of FeSe in the BCS-BEC cross-over regime for an approximately constant value of the magnetic field~\cite{kasahara.watashige.14}.
This result is very interesting due the fact that relatively large Maki parameter $\alpha_{M}$~\cite{maki.66}, describing the ratio of the critical magnetic fields related with the orbital and diamagnetic effects, has been measured for these type of materials and has been estimated in the range $3$-$7$~\cite{kida.matsunaga.09,kasahara.watashige.14,lei.wang.12,audouard.duc.15}.
As consequence the orbital effects of magnetic fields are negligible in these systems.
Then, the external magnetic field can lead to unconventional behaviour, such like the emergence of a superconducting state with Cooper pairs with non-zero total momentum.
Such a state is called the Fulde-Ferrell-Larkin-Ovchinnikov (FFLO) phase~\cite{FF,LO}
This phase is a partially-polarized superconducting state~\cite{matsuda.shimahara.07} and could be stable in iron-based superconductors~\cite{ptok.crivelli.13}.
However, our main motivation to a study of the influence of the MLT on the stability of the unconventional superconducting phase is the better theoretical understanding of relevant physical systems in close relation with forefront experiments, mentioned above.

The behaviour can be even  more complex in the case of two (or more) bands, especially if the system is e.g. superconducting, because the correlation effects (in a general case, the interactions can be different in every band) can modify the band structure regardless of the magnetic field.
For one band system cases, one can indicate only one MLT limit corresponding to the maximal/minimal value of band energy.
This band energy condition is also true in the case of a bigger number of bands, while the number of MLTs can be bigger.
This can lead to unconventional properties of the system, e.g. only FS for one type of spin survives.
Consequently, if superconductivity can exist in the system, such a state need to be the FFLO phase due to the fact that electrons in one band are fully spin polarized
while the partial spin polarization occurs in the second one.

\section*{Model and method}

First, let us describe the main idea of the MLT in a single band system.
In the absence of magnetic field, there is only one FS with spin-degeneracy in the system.
When an external magnetic field is applied, the densities of states are different for the electrons with spin down and spin up
and the population imbalance introduces a mismatch between the Fermi surfaces.
Hence, there are two FSs in the system.
With increasing magnetic field, the FS for the minority spin species disappears while the other one still exists.
The MLT takes place at the critical value of magnetic field for which one of FSs vanishes in the system.

\begin{figure*}[t!]
\centering
\includegraphics[width=\linewidth,height=8cm,keepaspectratio]{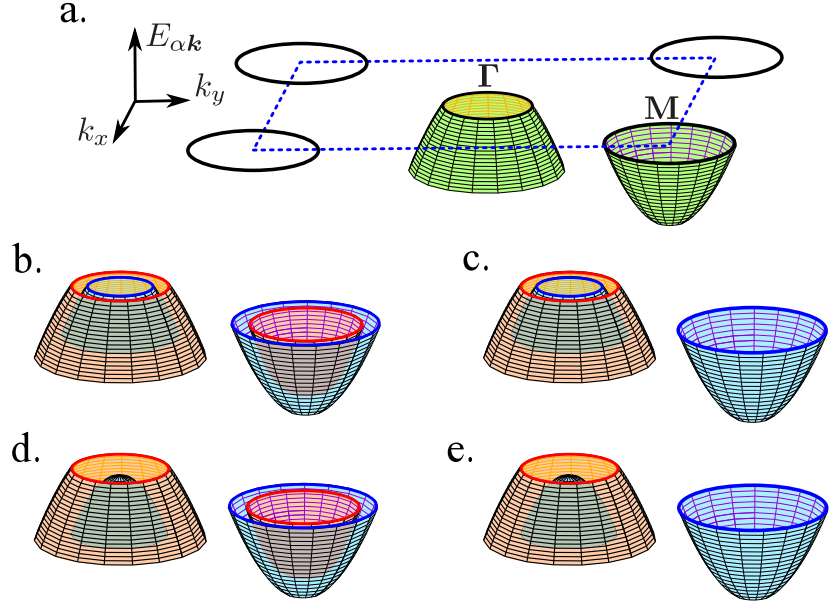}
\caption{ The main idea of the {\it magnetic Lifshitz transition} for a two-band system with hole- and electron- like bands, which are realised in the iron-based superconductors.
(a) The schematic band structure  and the Fermi surfaces (solid black lines) for iron-based superconductors in the absence of an external magnetic field in the first Brillouin zone with its boundaries indicated by dashed blue lines.
In the $\Gamma$ ($M$) point of the Brillouin zone, hole-like (electron-like) bands are located.
The three-dimensional plots show energy $E$ vs. 2D momentum ${\bm k}=(k_x,k_y)$.
(b) The application of the relatively small magnetic field leads to the splitting of the FSs in both bands.
Panels (c)-(e) show the band structure and the Fermi surfaces in the presence of an external magnetic field above the first (c) and (d), and the second (e), magnetic Lifshitz transition.
In all cases, the solid line represents the Fermi surfaces, while the blue (red) color corresponds to electron states with spin (anti)parallel to the magnetic field.
\label{fig.main}
}
\end{figure*}

The situation can be more complicated in a case of a multi-band system.
Now, we will discuss qualitatively the MLT in a two-band case, using a band structure similar to the iron-based band structures, in which one electron-like and one hole-like bands can be observed (left and right band respectively in each panel of Fig.~\ref{fig.main}).
In the absence of an external magnetic field, there are two FSs with double-spin-degeneracy (solid dark line in panel a of Fig.~\ref{fig.main}).
The application of the relatively small magnetic field leads to the splitting of the FSs in both bands.
As a consequence, four FSs, corresponding to each spin component, can be observed in the system (panel b).
However, for higher values of the magnetic field, one can find the MLT, which leads to the vanishing of one FS (panel c or d).
Further increase of the magnetic field can lead to the second MLT (panel e).
For the magnetic field values between these two MLT, we can find three FSs at Fermi level, while above the second MTL only two FSs
appear for partly filled subbands.
The transition from four to two FSs always occurs in a sequence of situations shown in panels b$\rightarrow$c$\rightarrow$e or b$\rightarrow$d$\rightarrow$e with increasing field and the particular sequence (one of both mentioned) is a consequence of the relation between widths of both bands.
Only for equal band widths, we can expect the transition b$\rightarrow$e, while the MLT arise in both bands at the same field.

To describe the characteristic band structure of iron-based superconductors we introduce the effective two-band model without hybridization ($t_{2g}$ orbital favour is conserved due to the orbital symmetry).
The non-interacting Hamiltonian $H_K$ considered here has the following form:
\begin{eqnarray}\label{eq:tuba}
H_{K} = \sum_{\alpha{\bm k}\sigma} ( E_{\alpha{\bm k}} - \mu ) c_{\alpha{\bm k}\sigma}^{\dagger} c_{\alpha{\bm k}\sigma}.
\end{eqnarray}
Here, $c_{\alpha{\bm k}\sigma}^{\dagger}$ ($c_{\alpha{\bm k}\sigma}$) denotes creation (annihilation) electron operators with spin $\sigma$ and two-dimensional (2D) momentum ${\bm k}$ in band $\alpha$, where $E_{1{\bm k}}$ ($E_{2{\bm k}}$) describe the hole-like (electron-like) band dispersion around $\Gamma$ ($M$) point in the FBZ, respectively.
Total number of particles in the system is given by the chemical potential $\mu$, while the influence of the external magnetic field $h$ is expressed by the Zeeman term:
\begin{eqnarray}
H_{H} = h \sum_{\alpha{\bm k}} \left( c_{\alpha{\bm k}\uparrow}^{\dagger} c_{\alpha{\bm k}\uparrow} - c_{\alpha{\bm k}\downarrow}^{\dagger} c_{\alpha{\bm k}\downarrow} \right) .
\end{eqnarray}
Because the FSs of iron-based superconductors have a cylindric shape~\cite{liu.zao.15}, such three-dimensional systems can be approximately described by a 2D model, which is equivalent to a small influence of the {\it z}-component of momentum on the band structure.
Notice that in iron-based superconductors the FS can be created by more than two bands, but top/bottom of these bands can be located far from Fermi level.
Thus, the two-band description of the MLT in these materials is sufficient, at least qualitatively.
For a sake of simplicity and without a loss of generality, we take the dispersion relations in the form:
\begin{eqnarray}
\nonumber E_{1{\bm k}} &=& - 2 t_{1} \left( \cos( k_{x}+\pi ) + \cos( k_{y}+\pi ) \right) - E_{1} , \\
\nonumber E_{2{\bm k}} &=& - 2 t_{2} \left( \cos( k_{x}+\pi ) + \cos( k_{y}+\pi ) \right) - E_{2} ,
\end{eqnarray}
where $E_{\alpha}$ is the shift of the center of $\alpha$ band ($\alpha=1,2$) with respect to the  chemical potential $\mu$ (the Fermi level in absence of interaction and magnetic field) determining the {\it band} filling ($n_{\alpha} = \sum_{\sigma} n_{\alpha\sigma}$), while $\mu$ defines the {\it global} filling of the system ($n = \sum_{\alpha} n_{\alpha}$).
Such a form of dispersion relations leads to hole-like behaviour for $-\mu-E_\alpha<0$ and to electron-like behaviour for $-\mu-E_\alpha>0$.
Here, $n_{\alpha\sigma}$ describes the average number of particles with spin $\sigma$ in band $\alpha$.
Because in the general case the band widths can be different, we take an additional parameter $\kappa = t_{2} / t_{1} > 0$ as the ratio between band widths in our model.
Moreover, we treat $t_{1}\equiv 1$ as the energy unit in the system.

In the following we assume that $\alpha=1$ band is hole-like band whereas $\alpha=2$ band is electron-like.
This assumption helps us to reproduce the effective band structure of iron-based superconductors
with hole-like and electron-like bands at $\Gamma$ and $M$ points of the FBZ, respectively (a case shown in Fig.~\ref{fig.main}.a).
$\kappa$ parameter describes the relation between the effective hole and electron effective masses (and it helps in the realization of the situation shown in Fig.~\ref{fig.main} (panels c and d)).
Moreover, adjusting $\kappa$ parameter leads to consideration of a model with a band structure corresponding to real-material band structure reported in ARPES experiment, where bottom (top) of the electron (hole) band are closer to the Fermi level than that of the second band \cite{kordyuk.12}.
This can correspond to band structure of real systems, such as FeSe system~\cite{kasahara.watashige.14} (for $\kappa < 1$) or KFe$_{2}$As$_{2}$~\cite{terashima.sekiba.09} (for $\kappa > 1$).

In the absence of coupling or at extremally weak coupling between bands, the interaction term can be written separately in the superconducting state for both bands~\cite{barzykin.09,hirschfeld.korshunov.11}, in the phenomenological BCS-like form:
\begin{equation}
\label{eg:ham.sc}
H_{SC} = \sum_{\alpha{\bm k}} U_{\alpha} \left( \Delta_{\alpha}^{\ast} ( {\bm k} ) c_{\alpha,-{\bm k}\downarrow} c_{\alpha{\bm k}\uparrow} + H.c. \right) ,
\end{equation}
where $U_\alpha$ is intra-band pairing interaction in band $\alpha$.
In our analyses we do not consider a possibility of an occurrence of the the FFLO phase, what is a consequence of an existence of only intra-band pairing \cite{ptok.crivelli.13,ptok.crivelli.15}.
All other possible interactions between particles are neglected \cite{nic.11}.
If inter-site pairing had existed the FFLO could be expected to occur in an absence of magnetic field.
There is no experimental evidences for such a behaviour.
Here, $\Delta_{\alpha} ({\bm k}) = \Delta_{\alpha} \eta_{\alpha} ( {\bm k} )$ denotes the superconducting order parameter (SOP) in bands $\alpha = 1,2$, respectively, and $\Delta_{\alpha}$ is an amplitude of the SOP.
$\eta_{\alpha} ( {\bm k} )$ is a form factor~\cite{ptok.crivelli.13,ptok.crivelli.15}, which describes a symmetry of the SOP and, for example, it equals  $1$ for {\it s-wave} symmetry or $\cos k_{x} - \cos k_{y}$ for {\it d-wave} symmetry.
This picture describes the intra-band pairing, while it can also be shown~\cite{ptok.14} that such a formulation describe an existence of the inter- and intra-orbital pairing in the system.
One should notice that in this paper formally two independent bands are considered, but because we treat both bands as one (two-band) system, they are coupled by the chemical potential $\mu$ and constant value of filling $n$.

\section*{Numerical results}

\begin{table}[b!]
\centering
\begin{tabular}{|c|c|c|c|}
\hline
Set label & $\kappa$ & $E_1/t_1$ & $E_2/t_2$ \\
\hline
$\mathcal{A}$ & 0.70 & 3.40 & -2.38 \\
\hline
$\mathcal{B}$ & 1.30 & 3.40 & -4.42 \\
\hline
\end{tabular}
\caption{\label{tab:parameters}
Model parameters $\kappa=t_2/t_1$, $E_1/t_1$, $E_2/t_2$ in two sets $\mathcal{A}$ and $\mathcal{B}$, which correspond to e.g. FeSe~\cite{kasahara.watashige.14} and Co-doped BaFe$_{2}$As$_{2}$~\cite{terashima.sekiba.09} compounds, respectively.
These sets correspond to cases, where the top (bottom) of the hole-like (electron-like) band is closer to the Fermi level.
$t_{1}$ is treated as the energy unit in the system.
In a real system, the bandwidths ($8t_\alpha$) can be taken approximately as $2$ eV~\cite{liu.zao.15}.}
\end{table}

The results for a given (global) filling $n$ are obtained by mapping of the ground state which is found by the minimization of the grand canonical potential, with respect to the SOP amplitude $\Delta_{0}$ for a fixed value of an external magnetic field $h$ and the chemical potential $\mu$, using the procedure described in Ref.~\citeonline{januszewski.ptok.15}.
Numerical calculations have been performed on the square lattice with $N_{x} \times N_{y} = 200 \times 200$ sites, what makes the finite-size effects negligible~\cite{ptok.crivelli.16}.
The additional parameters $E_{\alpha}$ are found from the following conditions: $n = 2$ (corresponding to $\mu/t_1=0$) and $n-n_{1} = n_{2} = 0.1$, in an absence of superconductivity (i.e., for the normal state).
It corresponds to the situation of almost fully-filled hole-like band and almost empty electron-like band.
We choose two sets of parameters $\{\kappa , E_{1}/t_1 , E_{2}/t_1\}$ listed in Table~\ref{tab:parameters}, which we will denote as $\mathcal{A}\equiv(0.70,3.40,-2.38)$ and $\mathcal{B}\equiv(1.30,3.40,-4.42)$ in the following sections of the paper.

To study the MLT we can define the function $\mathcal{F}_{\alpha\sigma} ( \omega )$:
\begin{eqnarray}
\mathcal{F}_{\alpha\sigma} ( \omega ) &=& \theta ( \omega + E_{\alpha\sigma}^{min} ) -  \theta ( \omega + E_{\alpha\sigma}^{max} )
\end{eqnarray}
where $\theta ( \omega )$ is the Heaviside step function, while $E_{\alpha\sigma}^{min}$ ($E_{\alpha\sigma}^{max}$) is the minimum (maximum) value of energy in $\alpha$ band for the electrons with spin $\sigma$.
This function is equal $1$ for energy $\omega\in[E_{\alpha\sigma}^{min}, E_{\alpha\sigma}^{max}]$ and zero otherwise.
It provides information about the existence of the FSs for a given spin-band index ($\alpha\sigma$), while the derivative,
\begin{eqnarray}
\left|\frac{ d\mathcal{F}_{\alpha\sigma}}{d\omega} \right| =
\delta( \omega + E_{\alpha\sigma}^{min} ) + \delta( \omega + E_{\alpha\sigma}^{max} ),
\end{eqnarray}
gives the peaks of the energy $\omega$ at the MLT.
Thus, the total number of the FSs in a multi-band system, in a state with the Fermi level $E_{F}$ is given as:
\begin{eqnarray}
\hat{\mathcal{F}} ( E_{F} ) &=& \sum_{\alpha\sigma} \mathcal{F}_{\alpha\sigma} ( E_{F} ) ,
\end{eqnarray}
while a peak of $d\hat{\mathcal{F}} / d\omega$ at $\omega=E_F$ denotes the MLT.

\begin{figure}[t!]
\centering
\includegraphics[width=\linewidth,height=5cm,keepaspectratio]{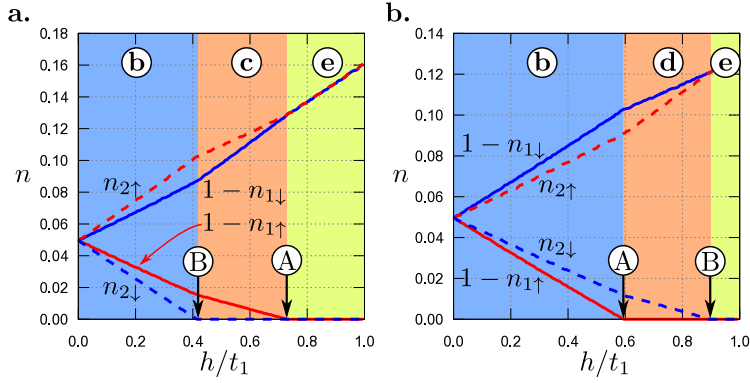}
\caption{
The average electron number $n_{\alpha\sigma}$ with spin $\sigma$ in band $\alpha$  as a function of external magnetic field $h/t_1$ (in the absence of pairing interaction, $U_\alpha=0$).
Results shown in panels a and b correspond to sets $\mathcal{A}$ and $\mathcal{B}$ of the model parameters, respectively.
Solid (dashed) lines correspond to 1st (2nd) band and red (blue) color is associated with spin $\uparrow$ ($\downarrow$).
The arrows show locations of two MLTs -- A (B) in circle indicates the place, where the 1st $\uparrow$-band (2nd $\downarrow$-band) becomes fully-filled (fully-empty).
The regions marked by small letters in circle, respectively, correspond to the schematic representation of the band structure in panels of Fig.~\ref{fig.main}.
\label{fig.lczno}
}
\end{figure}

In our case, in the absence of the interaction, the MLT limit is defined by the  state with spin $\downarrow$ and maximal energy in the first band and the state with spin $\uparrow$ and minimal energy in the second band, which can be changed by parameters $\kappa$ and $E_{\alpha}$ ($\alpha$=1,2).
It is important to emphasize that the value of $E_{F}$ in the system with a constant value of particle filling $n$ is a function of the magnetic field $h$. Moreover, $E_{F}$ can be also modified by correlation effects (e.g. by the superconducting state), what will be discussed in the paragraph below.

In the absence of interactions, the number of free electrons with spin $\sigma$ in band $\alpha$ is given by the simple formula,
\begin{eqnarray}
n_{\alpha\sigma}=\frac{1}{N}\sum_{\bm k} f(E_{\alpha{\bm k}}-E_{\alpha}-(\mu+\sigma h)),
\label{n_as}
\end{eqnarray}
where $f ( E ) = 1 / ( 1 + \exp ( - E / k_{B} T ) )$ is the Fermi-Dirac distribution.
The electron densities $\{n_{\alpha\sigma}\}$ (\ref{n_as}) are displayed in Fig.~\ref{fig.lczno}.
With increasing magnetic field, the number of particles with parallel spin $\uparrow$ increases in both bands (red lines).
Simultaneously, the number of particles with anti-parallel spin $\downarrow$ decreases (blue line).
At relatively large magnetic field, the number of particles with spin $\uparrow$ increases in the first (hole-like) band and the band becomes fully filled (arrow A), while the number of particles with spin $\downarrow$ decreases and the second (electron-like) band is fully empty (arrow B).
In such a case, two MLTs occur (denoted by arrows A and B).
As one can notice, the sequences of MLTs depends on the ratio $\kappa$.
The MLT at the narrower band  occurs in lower field than the MLT in the other band.
Note that the MLT in $\alpha=1$ ($2$) band is related with this subband becoming completely filled by (empty for) electrons with spin $\downarrow$ ($\uparrow$) at the transition.

\begin{figure}[t!]
\centering
\includegraphics[width=\linewidth,height=15cm,keepaspectratio]{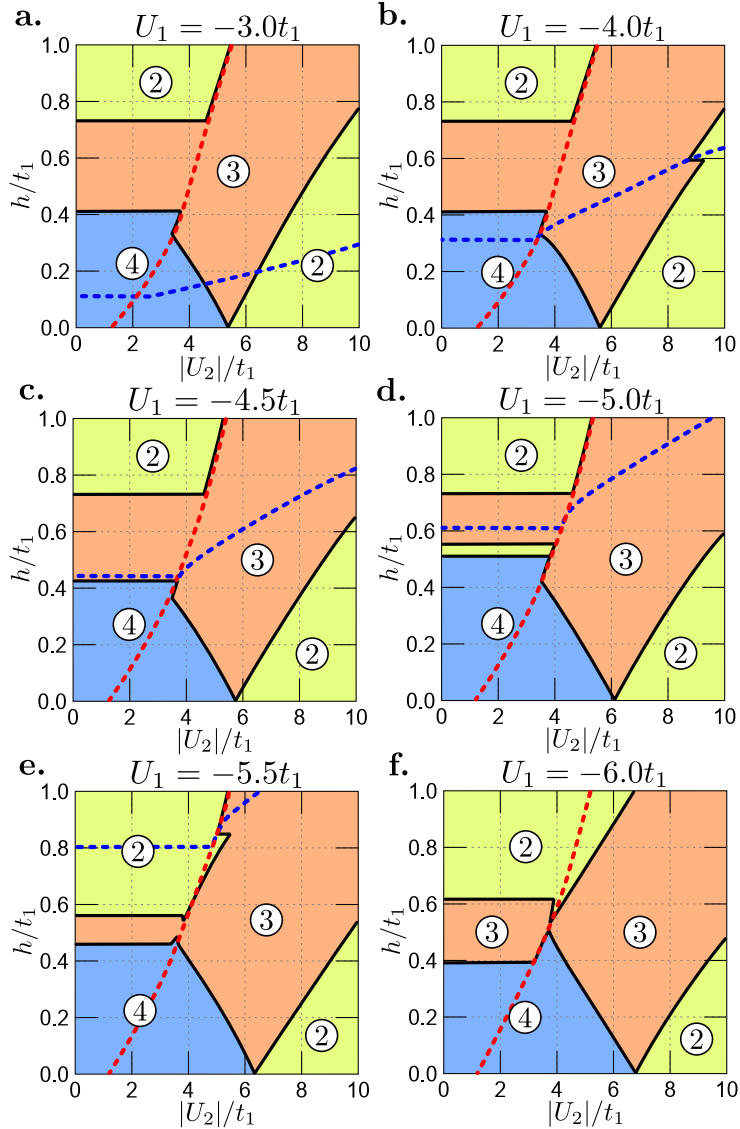}
\caption{
Magnetic field $h/t_1$ vs. $|U_2|/t_1$ phase diagrams for set $\mathcal{A}$ of the model parameters (Table~\ref{tab:parameters}) and different values of $U_1$.
Black solid lines denote the magnetic Lifshitz transition and numbers in circles denote numbers of subbands contributing to the Fermi surface in the system.
Blue and red dashed lines indicate upper critical field for 1st and 2nd band, respectively.
On panel f. upper critical field for the 1st band is located above the range of $h/t_1$-axis.
All transitions related to the disappearance of superconductivity (dashed lines) are discontinuous.
\label{fig.mlt.sc1st}
}
\end{figure}

\begin{figure}[t!]
\centering
\includegraphics[width=\linewidth,height=10cm,keepaspectratio]{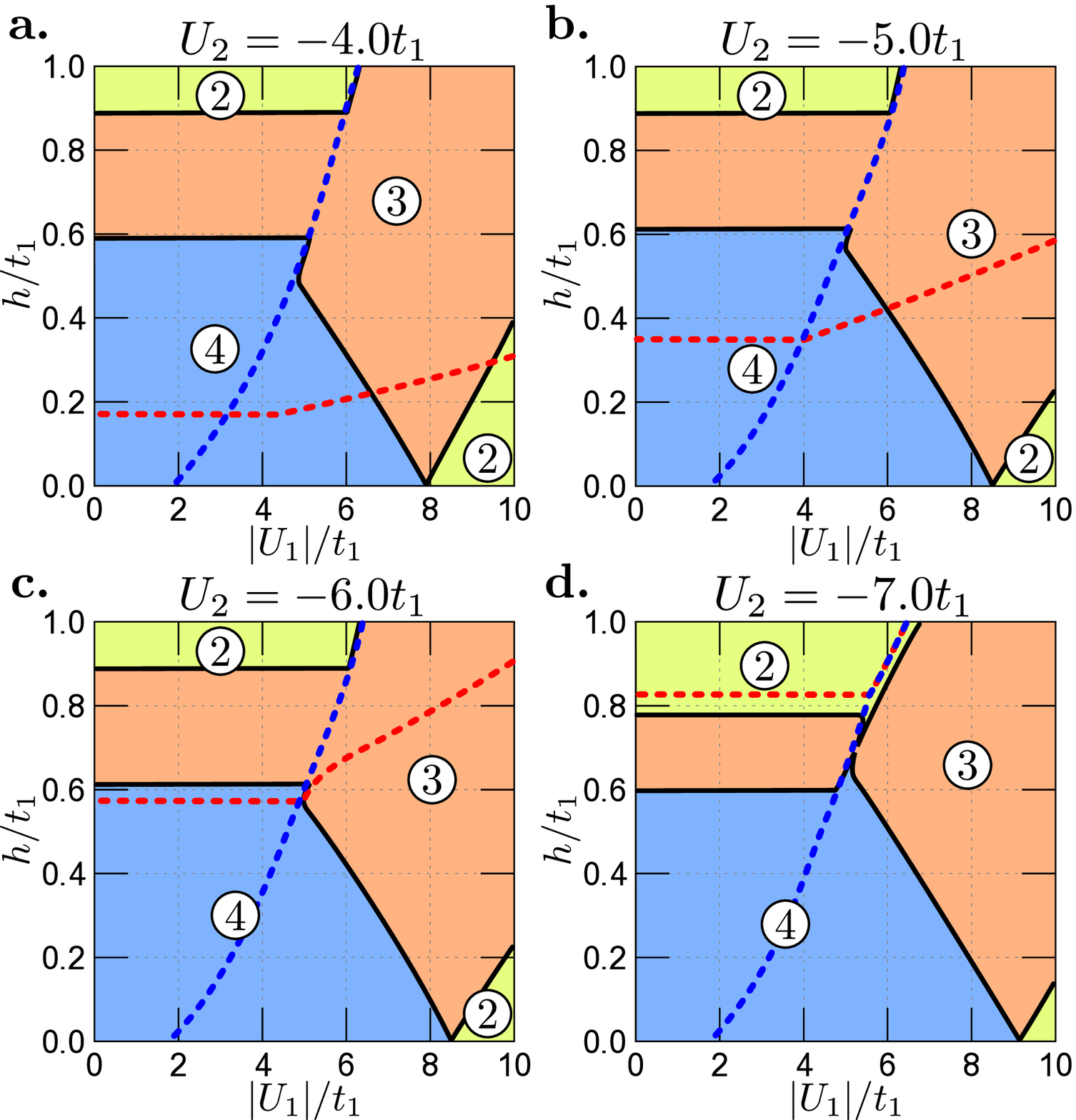}
\caption{
Magnetic field $h/t_1$ vs. $|U_1|/t_1$ phase diagrams for set $\mathcal{B}$ of the model parameters (Table~\ref{tab:parameters}) and different values of $U_2$.
Colour conventions for regions with $2,3,4$ FSs (numbers in circles) and the lines are the same as in Fig.~\ref{fig.mlt.sc1st}.
All transitions related to the disappearance of superconductivity (dashed lines) are discontinuous.
\label{fig.mlt.sc2st}
}
\end{figure}

As mentioned above, the Fermi level can be also modified by electron interactions.
For the present two-band model (\ref{eq:tuba})
the chemical potential $\mu$ depends on $U_\alpha$ and is calculated from global condition $n=2$.
Notice that filling in the bands ($n_1$ and $n_2$) can also be changed by interactions
$U_\alpha$ ($\alpha=1,2$) and they can differ from the values in the normal state.
The consequence of this is a change of the MLT limit by $U_\alpha$.
To keep this analysis general we assume different pairing interactions $U_{\alpha}$ in each band.
Thus, the superconductivity can vanish in each band at different magnetic fields.
In other words, the system exhibits two different critical magnetic field $h^{c}_{\alpha}$ which are defined as the magnetic field, where SOP in band $\alpha$ becomes equal zero.
However, critical magnetic field of the whole system is given as a bigger value from the set $\{h^{c}_{1},h^{c}_{2}\}$.
The numerical results for different values of $U_{\alpha}$ and previously chosen parameter sets $\mathcal{A}$ and $\mathcal{B}$ are shown in Figs.~\ref{fig.mlt.sc1st} and~\ref{fig.mlt.sc2st}, respectively.

Let us describe first the results for set $\mathcal{A}$ of the model parameters (Table~\ref{tab:parameters}) which are shown in Fig.~\ref{fig.mlt.sc1st}.
In the absence of the magnetic field ($h=0$), relatively large interaction $U_2$  leads to the standard LT, where the FS related to one of two-spin-degenerated bands disappears.
In the presence of the magnetic field, two independent MLT occur on the phase diagram (for small $h$, the lower
part of all panels). With increasing $h$ the distance between these two MLT increases (the characteristic V-shape on the phase diagrams for weak $h$).
For weak interaction $U_2$, e.g. in non-superconducting phase (left side of diagrams), two MLTs occur for increasing $h$.
The fields at the MLTs are independent of $U_2$ (black horizontal lines in panels a-c).
The occurrence of the superconductivity in the system (in one or both bands) leads to changes of the value of magnetic field at which the MLTs take place.
It is clearly seen that the MLT occurs in the neighbourhood of the upper critical field (solid dashed lines, e.g. panel e).
In the central part of the phase diagrams (i.e., in the presence of strong interaction), we can observe mutual influence of the magnetic field and electron pairing interactions
on the MLT.
As its consequence, the MLT line strongly depends on the parameters of the model (compare with e.g. Fig.~\ref{fig.mlt.sc2st}.d).
Notice that all transitions related to the disappearance of superconductivity at each band are associated with a discontinuous change of the SOP in the corresponding band.
As a consequence of that fact we can observe sharp  corners of the MLT lines occurring in the neighbourhood of these discontinuous transitions.

One observes that the results for set $\mathcal{B}$ of the model parameters, see Table~\ref{tab:parameters}, shown in Fig.~\ref{fig.mlt.sc2st} are qualitatively the same. In that case the electron-like band (i.e., the second band $\alpha=2$) is wider.
Notice that in Fig.~\ref{fig.mlt.sc2st} the parameters $U_1$ and $U_2$ are interchanged and values of $U_1/t$ are on the horizontal axes of the diagrams.

\begin{figure}[t!]
\centering
\includegraphics[width=\linewidth,height=10cm,keepaspectratio]{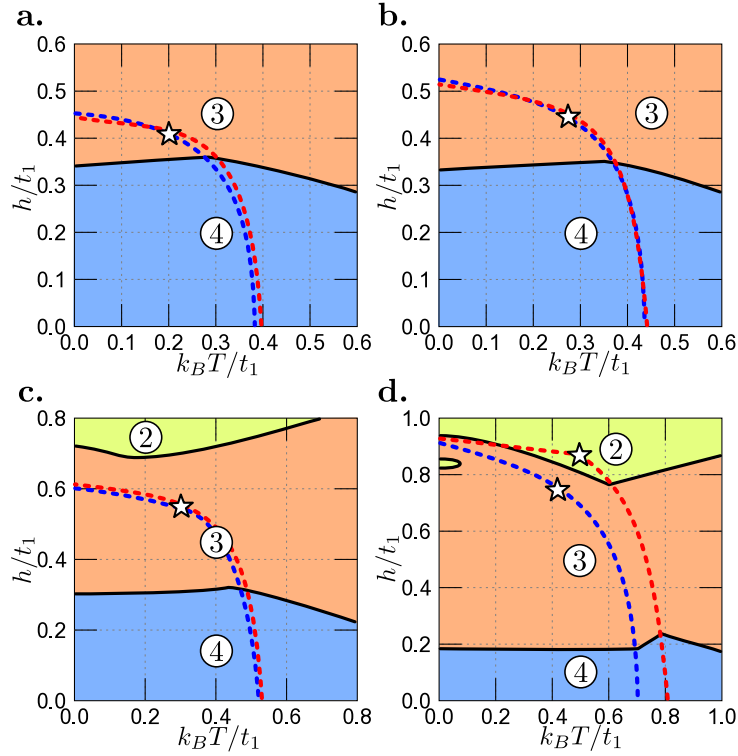}
\caption{
Magnetic field $h/t_1$ versus temperature $T$ phase diagram for set
$\mathcal{A}$ of the model parameters (see Table~\ref{tab:parameters}) and for several
values of pairing interactions ($|U_{1}|/t_{1}$, $|U_{2}|/t_{1}$):
($4.5,3.75$), ($4.75,3.95$), ($5.0,4.25$) and ($5.5,5.25$)
(panels a-d).
Black solid lines denote the magnetic Lifshitz transition and numbers
in circles stand for the number of partly filled subbands with Fermi
surface in the system.
Blue and red dashed lines indicate upper critical field for
the 1st and 2nd band, respectively.
At points denoted by star-symbols the transitions related to the disappearance of superconductivity  (dashed lines) change their type from first-order (at lower temperatures) into second-order (at higher temperatures).
\label{fig.htdiag}
}
\end{figure}

The most important problem related to the experimental results is the location of MLT on magnetic field $h$ versus temperature $T$ phase diagram.
Such diagrams for set $\mathcal{A}$ of the model parameters and fixed values of pairing interactions ($|U_1|/t_1$,$|U_2|/t_1$) are shown in Fig.~\ref{fig.htdiag}.
The choice of particular values of ($|U_1|/t_1$,$|U_2|/t_1$) in Fig.~\ref{fig.htdiag} are motivated by the existence of the same critical magnetic field for both bands.
Such a choice of parameters leads to the same critical temperatures for both bands approximately~\cite{ptok.14}.
For relatively weak interactions (panels a and b) one can find only one MLT in the superconducting phase.
This transition, associated with a change of the number of FSs in the system from $4$ to $3$,  exists near critical magnetic field.
Increasing the interactions (panels c and d) leads to the shift of that MLT to lower magnetic fields and an occurrence of second MLT slightly above (panel c) or in the superconducting region itself (panel d).
The second MLT transition is associated with a change of the number of FSs in the system from $3$ to $2$.
It should be stressed that temperature has a relatively small influence on the location of the first MLT in the superconducting phase.
However, in the presence of strong interactions the second MLT can be strongly modified
due to the interplay of superconducting state, high magnetic field, and temperature effects.
One should remark that the transitions related to the disappearance of superconductivity change their order from first-order (at lower temperatures) into second-order (at higher temperatures) at the points denoted by star-symbols in Fig.~\ref{fig.htdiag}, located approximately at temperature equal to a half of the critical temperature for each band in an absence of magnetic field (also cf. Ref.~\citeonline{matsuda.shimahara.07}).

Notice that the choice of concentration $n_1$ and $n_2$ in both bands as well as total concentration $n$ were arbitrary and in the general case it can be different. If $n_1$ were chosen smaller, the magnetic field at which the MLT occurs would also be smaller.

\section*{Discussion and summary}

In this paper, we have studied the specific type of the Lifshitz transition induced by the external magnetic field, which we call the magnetic Lifshitz transition (MLT).
We have discussed the mutual enhancement of the MLT and superconductivity in a two-band system, using the effective description of the iron-based materials in the
two-band model with electron-like and hole-like bands.
The effective bands model is simplified in order to gain a transparent physical insight and to avoid complicated details which would not be relevant for the MLT.
A more complex orbital model of iron-based superconductors, including the hybridization and Hund's coupling, is possible but
it would complicate the description of the studied system, not giving an advantage to the understanding of the phenomenon.
Our results show that in this system, for the well defined model parameters, we can identify two MLTs. They depend strongly on the band widths ratio.
The limit of the MLT can be modified by the pairing interaction in both bands but there is small influence of the temperature on the MLT occurrence.
Notice that values of magnetic field at which MLTs occur depend on the distance between top/bottom of the bands and on the total electron number (Fermi level).

We have shown that the MLT can be realised in the systems, where bottom (top) of the electron (hole) band are placed near the Fermi level.
The model considered here reproduces characteristic band structure of the iron-based superconductors \cite{kordyuk.12}, where hole-like and electron-like bands form around the $\Gamma$ and $M$ points, respectively (in the first Brillouin zone).
The chosen sets of the model parameters in Table~\ref{tab:parameters} can be connected with FeSe~\cite{kasahara.watashige.14} (in case of set $\mathcal{A}$) or Co-doped BaFe$_{2}$As$_{2}$~\cite{terashima.sekiba.09} compounds (for set $\mathcal{B}$).
Moreover, our resulting magnetic field versus temperature phase diagrams show rather weak influence of the temperature on the first MLT in the superconducting regime.
This result can be explicitly connected with the experimental observations for FeSe, where at low temperatures and high magnetic fields the additional phase transition within the superconducting dome can be observed~\cite{kasahara.watashige.14}.
On the other hand, both classes of iron based superconductors (11 and 122 family) show properties typical for a realisation the Fulde-Ferrell-Larkin-Ovchinnikov phase~\cite{ptok.15}.
In this context, further studies concerning possibilities of the occurrence of the Fulde-Ferrell-Larkin-Ovchinnikov phase and the MLT (as well as the interplay of these two phenomena) are important.

It should be noted that Hartree-Fock mean field approximation used in this work in general case overestimates critical temperatures and can give an incorrect description of the phases with a long-range order.
However, the approximation gives at least qualitative description of the system in the ground state, even in the strong coupling limit \cite{micnas.ranninger.90}.
Our results are obtained for three dimensional system in which the cylindrical dispersion relation are present.
As a result, effective system consists of noninteracting two dimensional planes.
In such a two dimensional system the upper boundary for superconductivity is the Kosterlitz-Thouless temperature,
which is about $\approx 75$ \% of the critical temperatures obtained in the mean-field approximation (in the range of model parameters considered)~\cite{denteneer.93}.
In real layered materials between weak but finite coupling between planes the real value of the temperature, where superconductivity vanishes is located between these two limits.
However, it does not change the qualitative behaviour of the magnetic Lifshitz transition inside the region of superconducting phases occurrence, which is presented in this study.

In this paper, we have discussed a role of the pairing iterations on the magnetic Lifshitz transition.
Although, the interactions between orbitals cannot be tuned in  condensed matter systems, a realisation of MLT and investigation of its properties can be performed experimentally with ultracold atomic Fermi gases in optical lattices
\cite{kohl.moritz.05,ospelkaus.ospelkaus.06,dutta.gajda.15}.
However, in this context it is still an experimental challenge.

\bibliography{biblio3}

\section*{Acknowledgements}

This work was supported by Large Infrastructures for Research, Experimental Development and Innovations project ``IT4Innovations National Supercomputing Center --- LM2015070'' of the Czech Republic Ministry of Education, Youth and Sports.
The support by Narodowe Centrum Nauki (NCN, National Science Centre, Poland),
    Project No. 2016/20/S/ST3/00274 (A.P.)
and Project No. 2012/04/A/ST3/00331 (A.M.O. and P.P.)
is also kindly acknowledged.

\section*{Author contributions statement}

A.P. arranged the project and performed numerical calculation.
All authors analysed and discussed the results.
A.C. consulted the results in the context of the BCS-BEC crossover.
The first version of the manuscript was prepared by A.P. and K.J.K.
All authors reviewed and contributed to the final manuscript.

\section*{Additional information}

\textbf{Competing financial interests}:
The authors declare no competing financial interests.

\end{document}